\documentclass[aps,prl,superscriptaddress,twocolumn,nofootinbib,floatfix]{revtex4-1} 

\usepackage[final]{graphicx}    
\graphicspath{{figures/}}       

\usepackage{dcolumn} 
\usepackage{bm} 
\usepackage{amsmath}
\usepackage{amsfonts} 
\usepackage{amssymb}
\usepackage[utf8]{inputenc}
\usepackage{textcomp}
\usepackage[english]{babel}
\usepackage[usenames,dvipsnames]{xcolor}
\usepackage{courier}
\usepackage{soul}
\usepackage{listings}
\usepackage{color}

\definecolor{dkgreen}{rgb}{0,0.6,0}
\definecolor{gray}{rgb}{0.5,0.5,0.5}
\definecolor{mauve}{rgb}{0.58,0,0.82}

 \newcommand{\mbf}[1]{\mathbf{#1}}

 \newcommand{\expectsm}[1]{\langle#1\rangle}
\begin{document}
\title{Simulating nonlinear dynamics of collective spins via quantum 
measurement 
and feedback}

\author{Manuel H. Mu\~{n}oz-Arias}
\author{Pablo M. Poggi}
\affiliation{Center for Quantum Information and Control, CQuIC, Department of 
Physics and Astronomy, University of New Mexico, Albuquerque, New Mexico 87131, USA}
\author{Poul S. Jessen}
\affiliation{Center for Quantum Information and Control, CQuIC, College of Optical Sciences
and Department of Physics, University of Arizona, Tucson, AZ 85721, USA}
\author{Ivan H. Deutsch}
\affiliation{Center for Quantum Information and Control, CQuIC, Department of 
	Physics and Astronomy, University of New Mexico, Albuquerque, New Mexico 87131, USA}

\begin{abstract}
We study a method to simulate quantum many-body dynamics of spin ensembles using measurement-based feedback. By performing a weak collective measurement on a large ensemble of two-level quantum systems and applying global rotations conditioned on the measurement outcome, one can simulate the dynamics of a mean-field quantum kicked top, a standard paradigm of quantum chaos. We analytically show that there exists a 
regime in which individual quantum trajectories adequately recover the classical 
limit, and show the transition between noisy quantum dynamics to full 
deterministic chaos described by classical Lyapunov exponents. We also 
analyze the effects of decoherence, and show that the proposed scheme 
represents a robust method to explore the emergence of chaos from complex 
quantum dynamics in a realistic experimental platform based on an atom-light 
interface.
\end{abstract}

%\date{Octubre 17 del 2014}
%\pacs{03.65.Vf,67.85.-d}

\maketitle
%\tableofcontents

%\section{Introduction}
\label{sec:intro}
%\begin{figure}[!h]
% \centering{\includegraphics[width=0.42\textwidth]{protocol_scheme.pdf}}
%\caption{Protocol scheme. See main text.}
%\label{fig:protocol_scheme}
%\end{figure}

The increasing level of precision achieved in both control and measurement of 
microscopic systems over the past decades is paving the way for using quantum 
systems as powerful simulators. In this paradigm, precise manipulation of a 
quantum system leads to the effective engineering of a particular physical 
model from which, in principle, one could extract quantities of interest which 
might not be accessible from a simulation on a classical device. In recent years prototypes of quantum simulators have been demonstrated in a variety of 
platforms, including trapped ions \cite{Blatt2012,Smith2016,Zhang2017}, cold 
atoms 
\cite{Bloch2012,Lewenstein2012,Bernien2017,Keesling2019,Aidelsburger2013,
Simon2011}, superconducting qubits \cite{Barends2015}, and photonic systems 
\cite{Aspuru-Guzik2012}.\\
\indent Although it is unclear whether current devices can reliably simulate 
complex dynamics beyond the capabilities of classical computers~\cite{Hauke2012}, recent 
explorations of small scale quantum simulations have proven to be interesting by 
themselves. For instance, simulations of interacting models which are native to 
cold atom implementations have led to new theoretical insights for weak 
ergodicity breaking~\cite{Bernien2017,Turner2018,Ho2019}. 
Quantum simulators can also be used to explore fundamental questions such as the emergence of classical behavior from quantum mechanics, typically referred to as the quantum-to-classical transition. This topic has motivated a vast amount of theoretical studies on different aspects of the problem, including the origin of quantum interference damping \cite{Zurek1993,Zurek2003}, the emergence of classical reality \cite{Ollivier2004,Zurek2009} and the role of the observer \cite{Nassar2013,Jeong2014}, alongside some experimental realizations \cite{Amman1998,Fink2010,Gadway2013,Unden2019}.

A long standing question is how the quantum-to-classical transition occurs for chaotic systems \cite{Zurek1995, Habib1998, Greenbaum2007}, given that unitary dynamics of closed quantum systems do not display exponential sensitivity to initial conditions \cite{Korsch1981,Hogg1982}, and in the chaotic regime, quantum dynamics follows classical motion only up to a time that is logarithmic in the system size (i.e. the Ehrenfest time)  \cite{Berry1979}.

Chaos in quantum dynamics has been understood from a variety of perspectives \cite{Peres1984, Toda1987, Schack1996, GarciaMata2018,Chavez2019}. One compelling approach is the emergence of classical chaos in the macroscopic limit of quantum systems as seen in the quantum trajectories associated with a continuous measurement~\cite{Bhattacharya2000, Bhattacharya2003, Ghose2004, Ghose2003a, Ghose2005}. 
% \textcolor{teal}{Historically in this approach nonprojective measurements have been used as a tool to constraint the interference between spatially distant regions of phase space, extending the Ehrenfest time to, in principle, arbitrarily long time scales, allowing for the recovering of the correct chaotic classical dynamics out of the continuously monitored coherent quantum evolution}. 
Particularly, Bhattacharya {\em et al.}~\cite{Bhattacharya2000} showed that when the measurement is strong enough to keep a quantum wave packet localized along the classical trajectory, but weak enough that measurement backaction does not dominate over Hamiltonian dynamics, the quantum trajectories display the correct Lyapunov exponents. Here, the effect of (nonprojective) measurements prevents the emergence of interference between spatially distant regions of phase space, thus effectively extending the Ehrenfest time to arbitrarily long time scales. A practical path to direct experimental observation of quantum trajectories characterized by a positive Lyapunov exponents, however, remains an open challenge \cite{Ralph2017}.\\

\begin{figure}[!t]
 \centering{\includegraphics[width=0.49\textwidth]{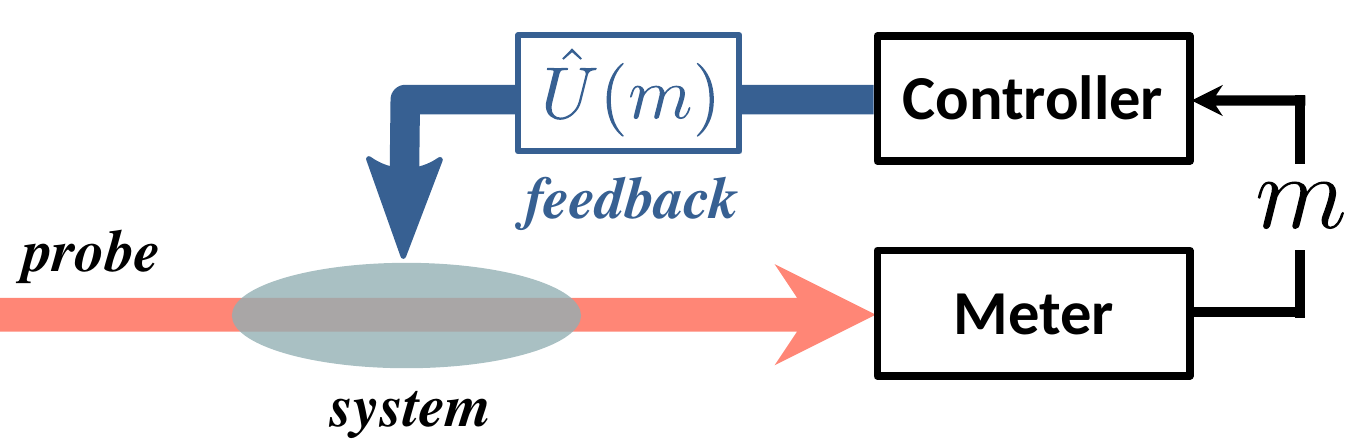}}
\caption{Schematic for a measurement-and-feedback based quantum simulator. A quantum system undergoes weak measurements, yielding an outcome $m$. This is then fed back to the system via a quantum control operation $\hat{U}(m)$.}
\label{fig:protocol_scheme}
\end{figure}

\indent In this Letter we propose a method to implement quantum simulations that are especially suited for exploring the emergence of chaos in quantum trajectories~\cite{Ralph2017,Pokharel2018,Eastman2019}. More generally, this proposal allows for the simulation of nonlinear dynamics in quantum systems described by collective spin variables, and thus constitutes a platform to explore complex phenomena such as phase transitions and criticality, \cite{Ivanov2019,Makhalov2019}. The method is based on performing a series of weak (nonprojective) measurements followed by global unitary control maps conditioned on the measurement outcome, see Fig. \ref{fig:protocol_scheme}. Our method is similar in spirit to the one originally proposed by Lloyd and Slotine~\cite{Lloyd2000}, who studied how measurement-based feedback could be used to simulate a novel form of chaos. Here we apply this procedure to study the simulation of mean-field dynamics which are chaotic in the thermodynamic limit. In particular, we simulate the kicked top (KT), a paradigmatic example of both classical and quantum 
chaos \cite{Haake1987,Schack1994,Constantoudis1997,Chaudhury2009}. We prove 
that the unconditioned dynamics generated by our protocol is a 
dephased version of the quantum KT (QKT), while the conditioned quantum
trajectories continuously approach the classical KT (CKT) dynamics as the size of the ensemble grows, and display the correct, positive largest Lyapunov exponent.  
%The proposed scheme is general and can be applied in different platforms. 
Finally, we explore a  potential experimental implementation based on an atom-light interface~\cite{Deutsch2010,Baragiola2014} and show that our results can be robust to the effects of decoherence.

\indent Consider an ensemble of $N$  noninteracting two-level systems described 
by collective spin operators $\hat{\bf{J}}=\frac{1}{2}\sum_{i=1}^N \hat{\bm{\sigma}}_i$, 
where $\hat{\bm{\sigma}}_i$ is a vector of Pauli matrices acting on the $i$-th 
particle. We take the system  initially  prepared in a spin coherent state (SCS), i.e., a product state of the form 
$\lvert \uparrow_{\vec{e}_\mbf{n}} \rangle^{\otimes N}$, where  $\vec{e}_{\bf{n}} \leftrightarrow 
(\theta,\phi)$ is an arbitrary direction on the unit sphere.  Each step of our protocol consists of two consecutive operations: (i) a nonprojective measurement of the $\hat{J}_z$ component of the collective spin, (ii) a unitary map conditioned on the measurement outcome. We consider measurements with Gaussian noise 
\cite{Jacobs2014}, yielding Kraus operators of the form
\begin{equation}
 \label{eqn:krauss_gauss_noise}
 \hat{K}_m = \frac{1}{(2\pi\sigma^2)^{1/4}} 
e^{-\frac{1}{4\sigma^2}\left(\hat{J}_z - m\right)^2}, 
\end{equation}
where $\sigma$ is the measurement resolution and $m$ is the measurement outcome, 
sampled with probability $P_m = 
{ _i\langle} \psi|\hat{K}_m^\dagger\hat{K}_m|\psi\rangle_i$. 
In each evolution step, the state is updated following 
quantum Bayes rule~\cite{Jacobs2010}, 
\begin{equation}
 \label{eqn:feedback_map}
 |\psi \rangle_{i+1} = \frac{\hat{K}_{\rm map}^{(m)} |\psi 
\rangle_i}{\sqrt{P_m}}, \enspace\text{where}\enspace \hat{K}^{(m)}_{\rm map} = 
\hat{U}\left(f(m)\right)\hat{K}_{m}.
\end{equation}\\
\noindent Here $\hat{U}\left(f(m)\right)$ is an unitary operator conditioned, via a 
feedback policy $f(m)$, on the measurement outcome.  $f(m)$ can be 
chosen with complete freedom, allowing the protocol to 
simulate different kinds of nonlinear dynamics in the collective spin variables. 
We will consider only global $\mathrm{SU(2)}$ rotations for our $f(m)$, since they can be implemented in state of the art experiments with relative ease~\cite{Chaudhury2009, Montano2018}. 
%\pablo{We point out here that the use of weak measurements and feedback was explored in \cite{Lloyd2000} for engineering nonlinear Schr\"odinger equations.}

% \textcolor{teal}{The use of feedback to simulate nonlinear dynamics, in the form of nonlinear Schrödinger equations, was first studied in}~\cite{Lloyd2000}.
%since these can be regarded as ``free" operations \cite{Albarelli2018}.\\
%We will explore complex nonlinear 
%dynamics of the collective spin system via the interplay of measurement and 
%conditioned rotations.
%\indent In order to show how this feedback protocol works, 

%Consider now the simulation of the KT dynamics. 
The dynamics of the QKT is governed by the Floquet operator \cite{Haake1987}
\begin{equation}
 \label{eqn:floquet_qkt}
 \hat{U}_{\rm QKT} = e^{ip \hat{J}_y} e^{i\frac{k}{2J}\hat{J}_z^2},
\end{equation}
describing the collective spin $J = N/2$ periodically undergoing a ``twist" around the $z$-axis characterized by strength $k$, followed by a rotation by angle $p$ about the $y$-axis. The classical limit of the Floquet map is obtained by considering the Heisenberg equations of motion in the limit $J \rightarrow \infty$, yielding a map for the vector $\mbf{n} \equiv \expectsm{\hat{\mbf{J}}}/J$~\cite{Haake1987}. The dynamics of the CKT is known to change from completely regular to fully chaotic as $k$ increases \cite{Haake1987}. Quantum signatures of this transition have been extensively studied in terms of hypersensitivity to perturbations 
\cite{Schack1994,Schack1996} and generation of entanglement \cite{Kumari2018,Kumari2019}.  
For the remainder of this work, unless otherwise stated, we will use $p=\pi/2$.

\begin{figure}[!t]
 \centering{\includegraphics[width=0.48\textwidth]{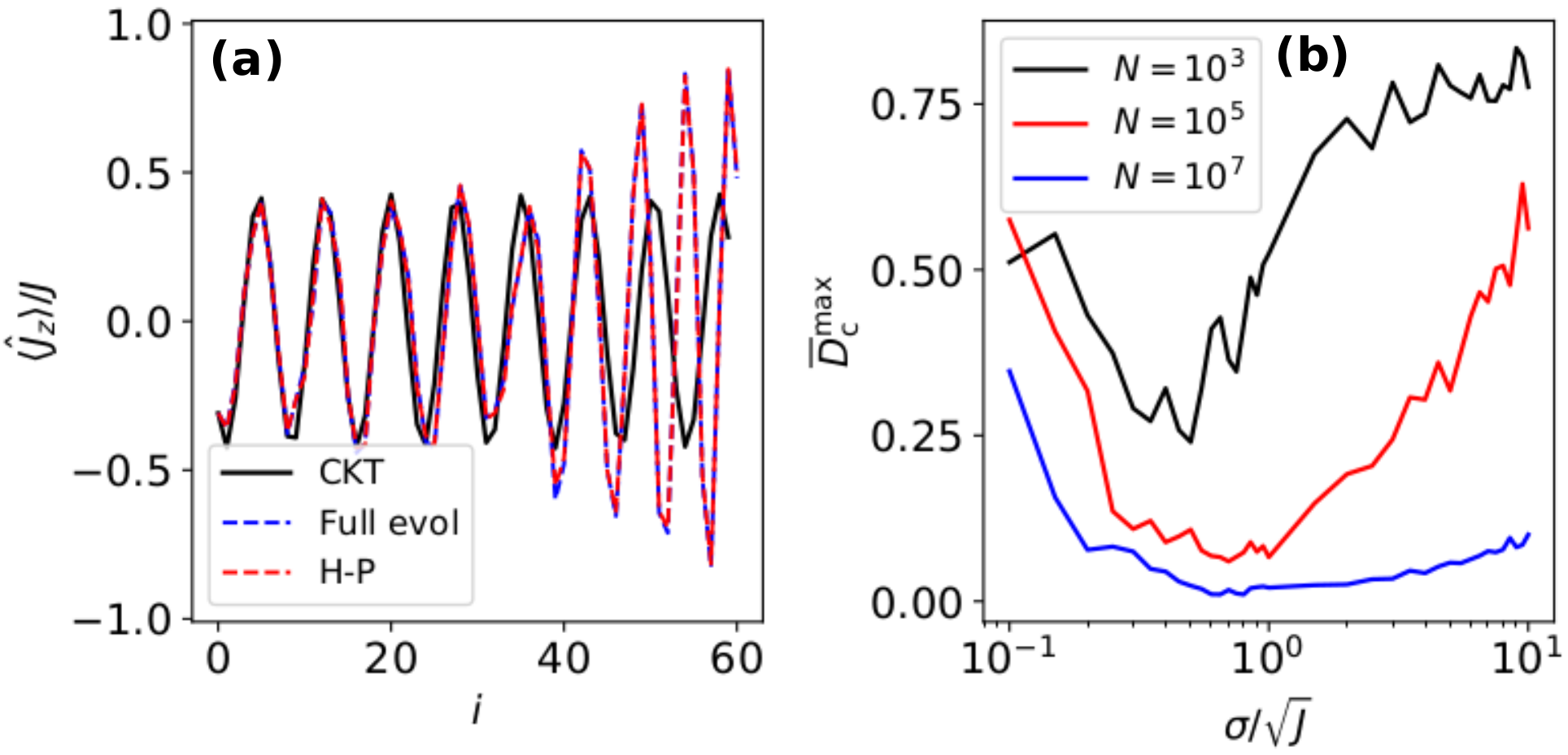}}
\caption{ (a) Time evolution of $\langle \hat{J}_z\rangle/J$ for regular dynamics. Dashed blue: full quantum evolution, dashed red: H-P approximation, black: classical KT. (b) Maximum distance to the corresponding classical trajectory averaged over several initial conditions $\overline{D}_{\rm c}^{\rm max}$ as a function of the measurement resolution. An optimal value of $\sigma$ exists around $\sigma \sim \sqrt{J}$. For both cases, parameters are: $k = 1.5$, $N = 10^3$, $\sigma = 0.9\sqrt{J}$.}
\label{fig:figure_1}
\end{figure}

To simulate the dynamics of the KT with our protocol, we note that the classical limit is equivalent to mean-field theory, where operators are replaced by their expected value, and correlations are neglected so that $k\hat{J}_z^2 \rightarrow 2k\langle\hat{J}_z\rangle\hat{J}_z$. We thus propose the feedback policy,
\begin{equation}
 \label{eqn:feedback_policy_kt}
 \hat{U}\left(f(m) \right) = \hat{U}_{p} \hat{U}_{k,m} = e^{ip\hat{J}_y} 
e^{i\frac{k}{J}m\hat{J}_z},
\end{equation}
which preserves the form of the free rotation and replaces the twisting 
unitary by a rotation conditioned on $m$. When $m = \expectsm{\hat{J}_z}$ we exactly recover the CKT. 
%Other models of collective spin dynamics can be accessed by changing the feedback policy.

%via a modification of the feedback policy, and a rich variety of nonlinear maps can be engineered by feeding back nonlinear functions of the measurement outcome.

\indent To see the connection of our feedback-based map to the QKT, we consider the evolution obtained by averaging over all possible outcomes 
\cite{Jacobs2014}. The state is now mixed, and the stroboscopic evolution of the density operator can be exactly obtained~\cite{SupMat}. The resulting map reads
\begin{equation}
\label{eqn:avg_map_1}
 \hat{\rho}_{i+1} = \sum_m \hat{K}_{\rm map}^{(m)} \hat{\rho}_i \hat{K}_{\rm 
map}^{(m)\dagger} = \hat{U}_{\rm QKT} \left( e^{\Gamma \mathcal{L}_D} [\hat{\rho}_i] \right) 
\hat{U}_{QKT}^\dagger.
\end{equation}

%The stroboscopic map for the density operator then {\color{blue} takes the form of the QKT with dephasing due to tracing over the measurement record,}
%\begin{equation}
%\label{eqn:avg_map_2}
%\rho_{i+1} = \hat{U}_{\rm QKT} \left( e^\mathcal{L} \rho_i \right) 
%\hat{U}_{QKT}^\dagger,
%\end{equation}
\noindent From Eq. (\ref{eqn:avg_map_1}) we see how the exact Floquet operator of the QKT emerges from our proposed scheme, in addition to dephasing generated by
\begin{equation}
\label{eqn:avg_map_3}
\mathcal{L}_D [\hat{\rho}_i] = - \left[ \hat{J}_z, \left[ 
\hat{J}_z, \hat{\rho}_i \right] \right],\ \mathrm{and}\  \Gamma = \frac{k^2\sigma^2}{2J^2} + \frac{1}{8\sigma^2}. 
\end{equation}
\noindent The dephasing rate $\Gamma$  arises from two effects: randomness in the measurement outcome and measurement backaction.  The first leads to randomness in the applied feedback, which increases with $\sigma^2$. The second decreases as $\sigma^{-2}$, as weaker measurement implies weaker backaction. The combination of both terms then has a minimum~\cite{SupMat}, yielding a regime of optimal tradeoff between information extraction and decoherence.

We now turn our attention to the conditioned dynamics of individual quantum trajectories. In this case, the state of the system depends on a series of measurement outcomes $m_1,m_2,...,m_n$ where $n$ is the number of time evolution steps. We will show that the simulated complex dynamics can be characterized via positive Lyapunov exponents~\cite{Lichtenberg1992,Wimberger}, a signature of chaos and unpredictability~\cite{Boffetta2002}. These exponents are associated with quantum trajectories, in the limit where the measurement acts to keep the state of the system sufficiently localized in phase space, and the effect of measurement backaction is negligible when compared to the applied unitary rotation~\cite{Bhattacharya2000,Bhattacharya2003}.

%For the collective spin systems under consideration, 
We are interested in working close to the classical limit, which is achieved for large spin ensembles with $N \gg 
1$, and for an optimal value of measurement resolution $\sigma$ allowing us to maximally extract an estimate of the mean field with minimal quantum backaction. In this limit, we can make the Holstein-Primakoff (H-P) approximation~\cite{Holstein1940}, and treat the state at all times as a Gaussian bosonic mode on a plane co-moving with the rotating Bloch vector. As shown in~\cite{SupMat}, this is an excellent approximation for the large ensembles under consideration here. In this approximation, the state is completely determined by the vector of mean values $\mbf{n}=\langle \hat{\mbf{J}} \rangle/J$ and a $3\times3$ symmetric covariance matrix $\mathbb{V}$ defined as
\begin{equation}
\mathbb{V}_{\alpha \beta} = \frac{1}{2J}\left(\langle \{\hat{J}_\alpha,\hat{J}_\beta\} \rangle - 2\langle{\hat{J}_\alpha}\rangle \langle{\hat{J}_\beta}\rangle\right),
\end{equation}
with $\alpha,\beta=x,y,z$. At each time step, a different H-P plane
is defined perpendicular to the direction of $\mbf{n}$. The state is expressed in the local basis on the plane, $\bm{n}'$ and $\mathbb{V}'$, via a rotation matrix, $\mathbb{A}$, taking $(\vec{e}_x,\vec{e}_y,\vec{e}_z)$ to $(\vec{e}_{\mbf{n}_1}, \vec{e}_{\mbf{n}_2}, \vec{e}_{\mbf{n}})$.  In this local basis the action of the Kraus operator only updates the subblocks of $\bm{n}'$ and $\mathbb{V}'$ corresponding to the two directions on the plane, while leaving the perpendicular one unchanged. The result is the familiar measurement-induced spin squeezing along the $P$ quadrature~\cite{Polzik2010}. The unitary feedback operation is trivially represented as a rotation of the H-P plane
conditioned on the measurement outcome. A particular example is shown 
in Fig. \ref{fig:figure_1}a where we compare the results for $\langle 
\hat{J}_z \rangle/J$ obtained with the full evolution of the state vector and 
the H-P approximation for $N = 10^3$. 

Using the H-P approximation, we can write down an analytic expression for the stroboscopic map evolving the normalized expectation values of the Cartesian components of $\mbf{n}$, which reads
\begin{subequations}
 \begin{align}
  \label{eqn:p_mean_vals_1}
 X_{i+1} &= -Z_i + \eta_1\mathbb{V}_{22}^{(i)}(1-Z_i)^2, \\
  \label{eqn:p_mean_vals_2}
 Y_{i+1} &= (1 - \eta_1\mathbb{V}_{22}^{(i)}Z_i)\left[ Y_i\cos\left(kZ_i + \eta_2\right) - 
X\sin\left(kZ + \eta_2\right)\right] \nonumber \\
  &+ \eta_1\mathbb{V}_{12}^{(i)} \left[X_i\cos\left( kZ_i +\eta_2 \right) + Y\sin\left( 
 kZ_i + \eta_2\right)\right], \\
  \label{eqn:p_mean_vals_3}
Z_{i+1} &= (1 - \eta_1\mathbb{V}_{22}^{(i)}Z_i)\left[ X\cos\left(kZ_i + \eta_2\right) + 
Y\sin\left(kZ_i + \eta_2\right)\right] \nonumber \\
 &- \eta_2\mathbb{V}_{12}^{(i)}\left[Y\cos\left(kZ_i + \eta_2\right) - 
X\sin\left(kZ_i + \eta_2\right)\right].
 \end{align}
\end{subequations}
The covariance matrix evolves similarly by a stochastic map.  The stochastic parts of these maps are characterized by
$\eta_1$ and $\eta_2$, which are normally distributed random variables with zero mean and 
variances given by
\begin{equation}
\sigma_1^2 = \frac{\sigma^2 + \Delta J_z^2}{\sigma^4},\ \mathrm{and}\ \sigma_2^2 
= k^2\frac{\sigma^2 +\Delta J_z^2}{J^2},
\end{equation}
where $\Delta J_z^2$ is the spin uncertainty (``projection noise").
From these expressions it is easy to see that these random corrections $\eta_1$ and $\eta_2$ vanish as  $J \to \infty$, and thus the map limits to the exact CKT dynamics. 
\begin{figure}[!t]
 \centering{\includegraphics[width=0.48\textwidth]{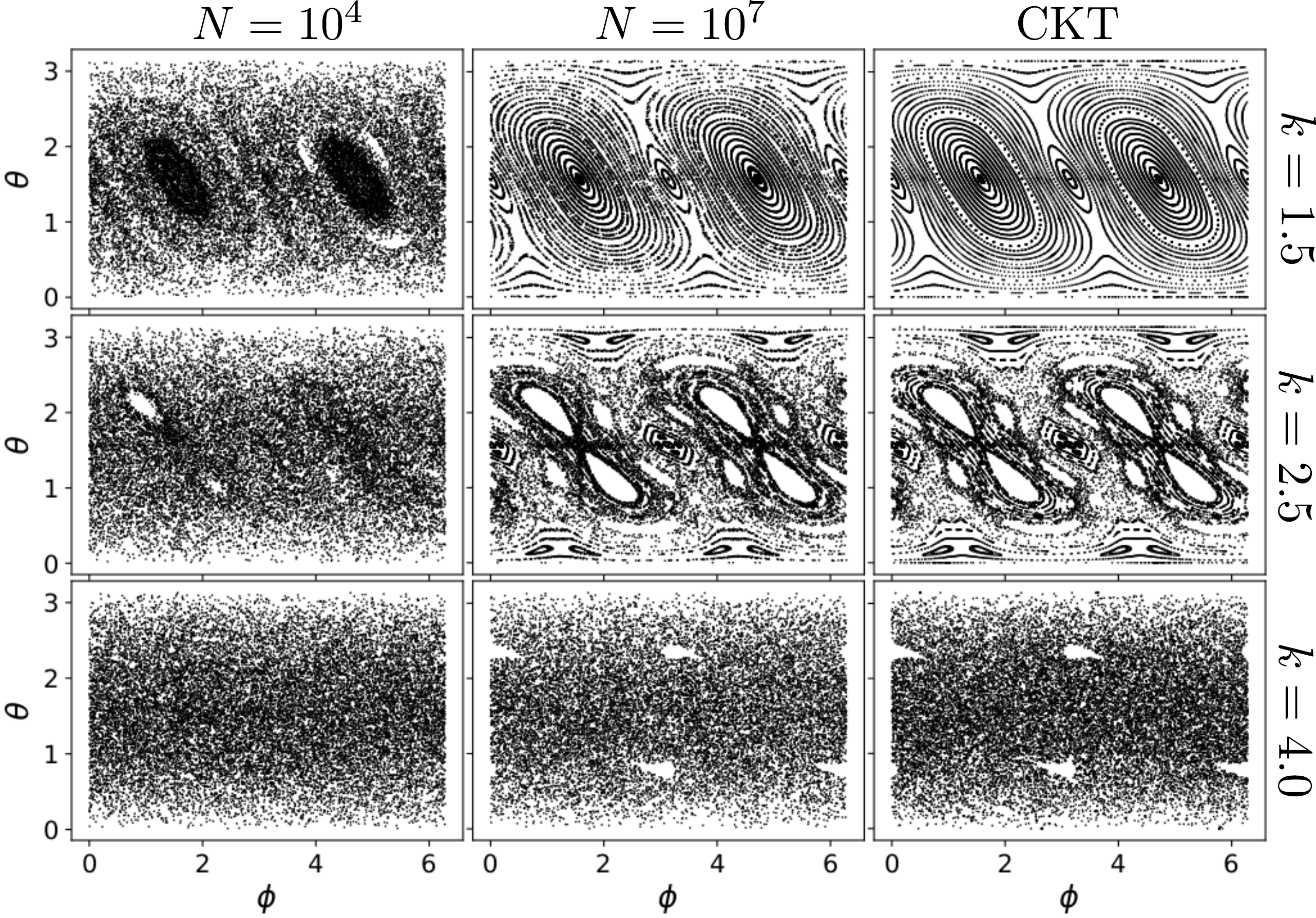}}
\caption{Phase space portraits constructed using the H-P approximation with $\sigma = 0.9\sqrt{J}$. From top to bottom we show $k = 1.5, 2.5, 4.0$, respectively. The first two columns correspond to $N = 10^4, 10^7$, the third column are the CKT portraits. The emergence of the classical regular, mixed and chaotic features can be seen as $N$ increases.}
\label{fig:figure_2}
\end{figure}

For a finite-sized system, the stochastic corrections $\eta_1$ and $\eta_2$ 
quantify the effects of noise introduced due to the measurement backaction and 
an imperfect feedback operation, respectively. As in the case of the 
average map, one can find the value of measurement resolution which minimizes 
these effects (see~\cite{SupMat}).  We get a good estimate of the mean value of $J_z$ when the shot noise resolution of the meter is on the order of the projection noise of the SCS. For the large ensembles of interest here, this corresponds roughly to $\sigma/\sqrt{J}\sim 1$. To illustrate this point, we computed the largest distance to the respective classical trajectory $D_{\rm c}^{\rm max} = \max\limits_i ||\bm{X}_{\rm CKT}^{(i)} - \bm{n}^{(i)}||$ and calculated its average over $100$ initial conditions 
for a regular phase space ($k=1.5$). The results are shown in Fig. 
\ref{fig:figure_1}b as a function of $\sigma/\sqrt{J}$. There, we observe the existence of an optimum regime which becomes less and less 
restrictive as the system size $N$ increases.

Using the co-moving H-P approximation we can model arbitrary ensemble sizes, 
and thus study extensively the phase spaces generated by our protocol with 
optimal measurement strength. Fig. \ref{fig:figure_2} shows the phase portraits for different strengths of the chaoticity parameter, $k$, and ensemble sizes. For small ensembles, quantum noise washes out the classical features. Regular, mixed and chaotic features of the CKT emerge in the large $N$ limit, becoming essentially indistinguishable from the classical phase portrait for {$N\geq 10^7$}.
% Hence, the individual quantum trajectories generated using this protocol simulate the classical KT dynamics, reproducing the regular, mixed and chaotic features of the phase space. The degree of accuracy of the simulation increases continuously as the system size increases. 

\begin{figure}[!t]
 \centering{\includegraphics[width=0.49\textwidth]{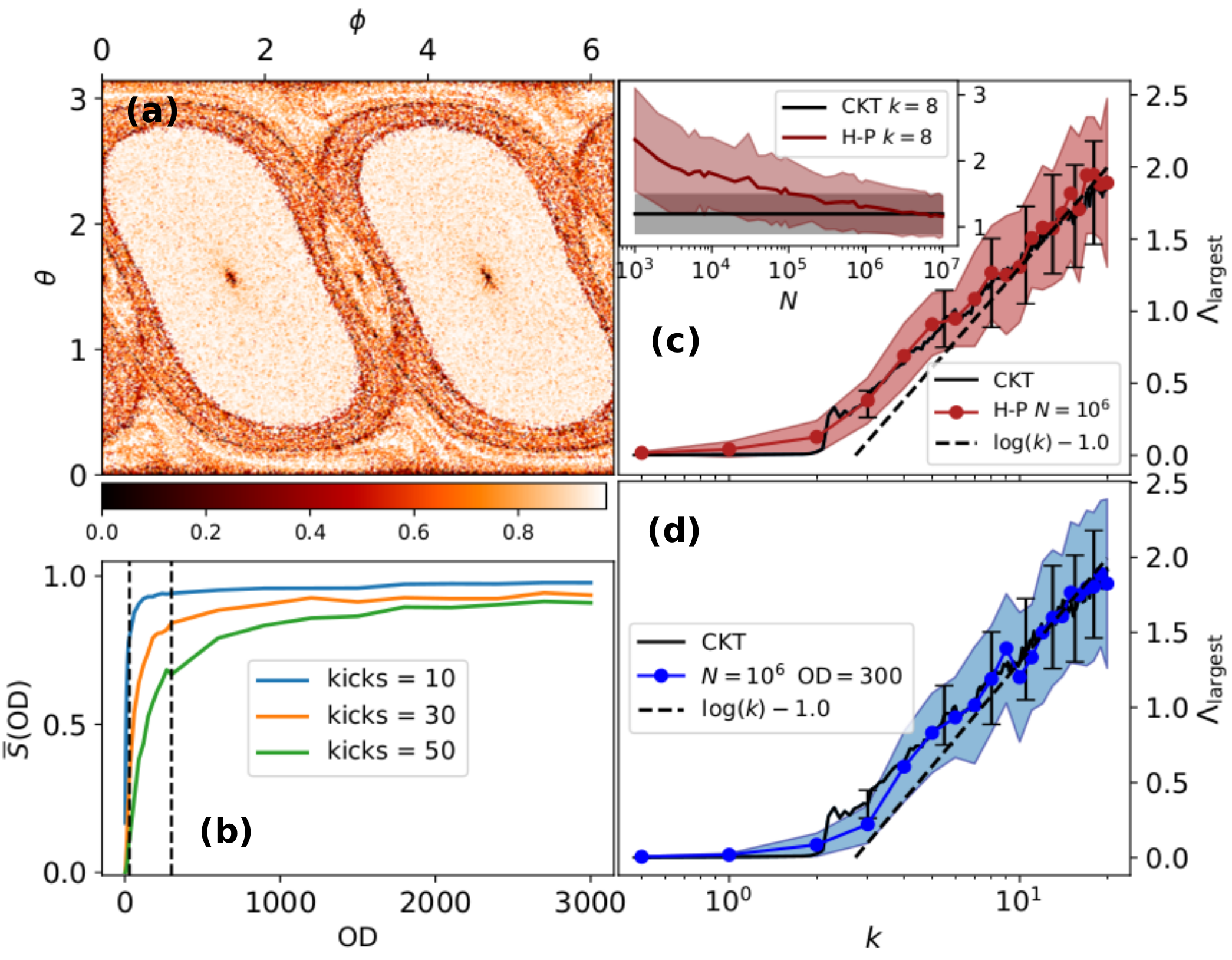}}
\caption{Robustness of simulated classical dynamics in the presence of the probe-induced decoherence.  characterized by the optical depth on resonance $OD$, which determines the cooperativity, \textit{i.e.} the measurement rate compared to the decoherence rate. \textbf{(a)} Point like comparison between the CKT phase portrait and the phase portrait of our model using $S({\rm OD}; \sigma, k, n)$ in Eq. (\ref{eqn:similarity}). With, $k = 1.5$, ${\rm OD = 300}$, $\sigma = 4.0\sqrt{J}$ and $n = 30$. \textbf{(b)} $S({\rm OD}; \sigma, k, n)$ averaged over initial conditions as a function of $\rm OD$. Vertical lines signal ${\rm OD = 30}$ and ${\rm OD = 300}$, respectively. Notice the large drop in similarity below ${\rm OD} \sim 100$. Parameters are as in \textbf{(a)}. \textbf{(c)} Largest Lyapunov exponent as a function of $k$. For the CKT (black continuous) and the H-P approximation (red dots). The inset shows the crossover from quantum to classical as seen in $\Lambda_{\rm Largest}$, as a function of $N$ and for $k = 8$. \textbf{(d)} Largest Lyapunov exponent in the presence of decoherence (blue dots). In \textbf{(c)} and \textbf{(d)} the dashed black line is an analytical expression known to work for $k>10$~\cite{Constantoudis1997}}.
\label{fig:lyaps_with_OP}
\end{figure}

We quantify the quantum-to-classical transition in both the chaotic and regular regime. For the regular case, as $N$ increases the maximum distance between the simulated trajectory and the classical trajectory approaches zero, see Fig. \ref{fig:figure_1}b. For the chaotic case, we study the convergence of the largest Lyapunov exponent, $\Lambda_{\rm Largest}$, and its statistical variance as a function of $N$ (see inset in Fig. \ref{fig:lyaps_with_OP}c). $\Lambda_{\rm Largest}$ is calculated as the average over the local exponents associated with initial conditions over the whole sphere.  Each of the local exponents is calculated as the average divergence rate between the fiducial initial condition and a set of nearby shadow initial conditions~\cite{Bhattacharya2000, SupMat}.   For values of $N$ for which our protocol agrees with the CKT,  Fig. \ref{fig:lyaps_with_OP}c shows  $\Lambda_{\rm Largest} $ obtained with our protocol as a function of of the twisting strength $k$. Taken together, the results demonstrate that, in the classical limit, our protocol simulates to a very good approximation both the regular and the chaotic dynamics of the CKT.

Finally, we consider a possible realization of our protocol based on an atom-light interface. A QND measurement of a projection of $\hat{\mbf{J}}$ can be implemented through the dispersive interaction between a laser probe and an ensemble of trapped laser-cooled atoms~\cite{Deutsch2010}. An example is the Faraday interaction, in which the polarization of an off-resonance laser probe rotates by an angle proportional to the collective spin magnetization along the direction of propagation. A subsequent measurement of the probe in a polarimeter provides a weak, continuous-time QND measurement $m(t)$ of the collective spin projection $\hat{J}_z$~\cite{Takahashi1999, Kuzmich2000}. Such QND measurements have been employed to create spin-squeezed states in a variety of experiments~\cite{Kuzmich2000,Takano2009,Sewell2012,Appel2009,Leroux2010,Schleier-Smith2010,Bohnet2014,Hosten2016}. Here we consider processing the measurement record in a classical controller, which can apply real-time feedback to the spin trough a set of magnetic coils to drive spin rotations around different axes, conditioned on $m(t)$ as desired.

For this scheme, we study how decoherence during the course of measurement affects our ability to observe the quantum-to-classical transition. %In this geometry, 
Measurement occurs at a rate at which photons are forward-scattered into the  probe, $\kappa=(\sigma_0/A)\gamma_s$ where $\sigma_0$ is the resonant scattering cross section, $A$ is the effective beam area, and $\gamma_s$ is the photon scattering rate into $4\pi$~\cite{Deutsch2010}. The measurement resolution variance $\sigma^2 =1/\kappa T$, where $T$ is the duration of the measurement. During this time, photons will be diffusively scattered, leading to optical pumping and concomitant decoherence. The duration of measurement is chosen so that $\sigma^2\approx \Delta J_z^2 \sim N$, which implies $\gamma_s T \sim 1/OD$, where $OD = N \sigma_0/A$ is the optical density, which plays the role of the cooperativity in the atom-light interface~\cite{Baragiola2014}.  Thus, for sufficiently large $OD$, we expect to be able to extract significant information with minimal decoherence as illustrated in Fig. \ref{fig:lyaps_with_OP}b.
% However, diffusive photon scattering events (spontaneous emission) lead to optical pumping (OP), limiting the performance of the measurement. When the atomic ensemble is optically thin at the detuning of the probe light, spontaneous emission between different atoms is uncorrelated. Thus, the effect of OP is described by a map acting locally on individual atomic spins. 

We study this by simulating the measurement record in a simplified model of the atom-light interface. The measurement outcome is given by the time average of the continuous measurement record, $m = \int_0^T dt \mathcal{M}(t)$, where
\begin{equation}
 \label{eqn:continuous_msmt_outcome}
 \mathcal{M}(t)dt = \mathrm{Tr}(\rho(t) \hat{J}_z ) dt + \frac{1}{\sqrt{\kappa}}dW,
\end{equation}
and $dW$ is a Wiener increment \cite{Jacobs2010}. The state of the atomic ensemble evolves according to a stochastic master equation~\cite{Jacobs2006,Jacobs2014}
%familiar in the description of continuous measurements~\cite{Jacobs2006,Jacobs2014},
\begin{equation}
d\rho = \frac{\sqrt{\kappa}}{2} \mathcal{H}[\rho] dW + \frac{\kappa}{8} \mathcal{L}_D[\rho]dt + \gamma_{\rm s} \sum_{i} \mathcal{D}_i[\rho] dt, 
\label{eqn:SME}
\end{equation}
where the map $\mathcal{H}[\rho_{\rm A}] = \{ \rho_{\rm A}, \hat{J}_z \} - 2\langle \hat{J}_z \rangle \rho_{\rm A}$ represents stochastic kicks and $\mathcal{L}_D$ represents dephasing, as in Eq. (\ref{eqn:avg_map_3}).  The additional map, $\mathcal{D}[\rho_{\rm A}]$,  accounts for optical pumping on individual atoms in the ensemble~\cite{SupMat}. 

Given the measurement record with decoherence, we calculate the similarity between phase spaces of the simulation and the CKT in the regular regime as follows. Let
\begin{equation}
\label{eqn:similarity}
S(OD; \sigma, k, n) = {\rm cor}(\bm{\theta}_{\rm CKT}, \bm{\theta}) {\rm cor}(\bm{\phi}_{\rm CKT}, \bm{\phi}) |\bm{n}|^2_{\rm min},
\end{equation}
where ${\rm cor}(\bm{A}, \bm{B})$ is the Pearson correlation coefficient of the vectors $\bm{A}$ and $\bm{B}$, $|\cdot|^2_{\rm min}$ is the minimum norm squared among all of the vectors $\bm{n}$ which compose our simulated trajectory, and $(\bm{\theta}_{\rm CKT}, \bm{\phi}_{\rm CKT})$, and $(\bm{\theta}, \bm{\phi})$ are the CKT and our model trajectories, respectively. With this measure, using several initial conditions, we construct a point to point similarity map between the CKT and our model with decoherence.  In Fig. \ref{fig:lyaps_with_OP}a we show results for $k = 1.5$, $n = 30$ and ${\rm OD} = 300$. We see a large portion of phase space which can be reproduced to a very high degree of accuracy. Interestingly, the regions of phase space which are harder to simulate correspond to fixed points and separatrix lines.

To study how decoherence affects our ability to observe the chaotic behavior we calculated $\Lambda_{\rm Largest}$ as a function of $k$ for $N \sim 10^6$, which fixes the value of $\rm OD=300$ (smaller $OD$ values are analyzed in ~\cite{SupMat}). In Fig. \ref{fig:lyaps_with_OP}d we observe good agreement with the CKT result, demonstrating that the protocol with modest decoherence is a robust scheme to explore the emergence of quantum trajectories characterized by the appropriate positive Lyapunov exponent.

%\section{Conclusions}
\label{sec:conclusions}
In summary, we proposed a measurement-based feedback 
protocol to simulate complex nonlinear dynamics in collective spin systems. 
For a well-chosen feedback policy, we showed that the average evolution gives 
rise to the one-axis twisting Floquet map of the QKT.  In the limit of large ensembles, individual quantum trajectories recover chaotic classical dynamics, 
characterized by the positive Lyapunov exponent of the 
CKT. Under a model implementation based on QND measurement in an atom-light interface and in the presence of decoherence, we explored conditions for which we can observe the quantum-to-classical transition. Our protocol opens the door to explorations of chaotic many-body dynamics~\cite{Prosen2018} and their implications for quantum simulations.

We thank Ezad Shojaee and Elizabeth Crosson for helpful discussions. This work was supported by NSF grants PHY-1606989, PHY-1607125, and PHY-1630114.

\bibliographystyle{apsrev4-1}
\bibliography{feedback_based}

\end{document}